\documentclass{ws-p10x7}

\begin{document}

\title{Hadronization of quark gluon plasma (and gluon jets) and the role of the 0++ glueball
as primary hadron products}

\author{ \underline{Peter Minkowski} }

\address{
 Institute for Theoretical Physics, University of Bern,
Sidlerstrasse 5, CH-3012 Bern, Switzerland
\\
E-mail: mink@itp.unibe.ch
}

\author{ Sonja Kabana}

\address{
 Laboratory for High Energy Physics, University of Bern,
Sidlerstrasse 5, CH-3012 Bern, Switzerland
\\
E-mail: sonja.kabana@cern.ch
}

\author{
Wolfgang Ochs
}

\address{
Max Planck Institut fuer Physik, Werner Heisenberg Institut,
Foehringer Ring 6, D-80805, Muenchen, Germany
\\ E-mail: wwo@mppmu.mpg.de }

\twocolumn[\maketitle\abstract{
\noindent
Signatures of dominant and central production of glueballs (binary gluonic mesons)
in heavy ion and hadromic collisions are discussed. Search strategies are 
proposed. 
}]

\section{Introduction}

\noindent
The three sequences of binary gluonic mesons (gb) are reviewed \cite{HFPM}, \cite{PMWO},
represented by the respective gb resonances with lowest mass and
$J^{PC}$ quantum numbers : $0^{++}$, $0^{-+}$ and $2^{++}$. 
While the $2^{++}$ sequence is associated through Regge analytic continuation in
angular momentum of two body elastic amplitudes to the Pomeron trajectory,
the triple Pomeron vertex is thought to be responsible for the
multiparticle production of mainly $0^{++}$ glueballs. We discuss search
strategies for the main $0^{++}$ component and also for the heavier $2^{++}$ state
in heavy ion and hadronic inelastic scattering at high energy and high (initial)
energy density under the hypothesis that they become the dominant primary
systems of multiparticle production in this environment \cite{SKPM}.  
We propose in particular to impose centrality selections also in
hadronic reactions, e.g. $p \overline{p}$ collisions at the Tevatron,
in order to discover eventual transitory behaviour bearing a similarity
to kaon distributions in Pb Pb collisions at the SPS as analyzed in ref. \cite{Sonja1},
\cite{Sonja2}.

\section{Discussion}

\noindent
The three $J^{PC}$ series of binary gluonic mesons are  

\begin{equation}
\begin{array}{l}
\label{eq:1}
\begin{array}{lll lll ll}
 & & & & & & &
\vspace*{0.3cm} \\
\hline
\vspace*{-0.2cm} \\
0^{++} & 0^{++} &   & 2^{++} &        & 4^{++} &        & \cdots
\vspace*{0.3cm} \\
2{++}  &        &   & 2^{++} & 3^{++} & 4^{++} & 5^{++} & \cdots
\vspace*{0.3cm} \\
0^{-+} & 0^{-+} &   & 2^{-+} &        & 4^{-+} &        & \cdots
\end{array}
\end{array}
\end{equation}

\noindent
We envisage two temporal developments of a collision :

\begin{description}
\item
a) quark gluon plasma formation

initial thermal equilibrium as appropriate for an expanding medium 
within the quark gluon plasma phase. In the further development the 
hadronisation process from within the plasma phase coincides closely in time 
with chemical freezout.  

\item
b) no quark gluon plasma formation

initial thermal equilibrium occurs within the hadronic phase with
subsequent chemical freezout.
\end{description}

\noindent
It is conceivable, that for a given centre of mass collision energy
both phases a) and b) above occur depending on the centrality of the collision.
Then phase a) is distinguished by the independence of the thermodynamic
intensive variables, mainly temperature and 
chemical potentials for baryon number and strangeness 
$T \ \sim \ T_{cr}$, $\mu_{b}$ and $\mu_{s}$ from the initial energy density,
prior to thermalization $\varepsilon_{\ 0}$ \cite{SKPMn}.

\noindent
It is for the case of phase a) that we expect dominant production
of $gb \ ( \ 0^{++} \ )$ to occur upon chemical freezout. We
discuss the possibility of characteristic $\pi^{+} \pi^{-}$ as well as $K \overline{K}$
invariant mass and relative momentum distributions to remain observable
notwithstanding further hadronic collisions before final thermal freezout.

\noindent
The following properties appear characteristic

\begin{description}
\item a1) dipion invariant mass distribution :

Ordering all pions produced in rapidity and relative momentum 
the invariant mass spectrum corresponding to $gb \ ( \ 0^{++} \ )$ production
should show the interference pattern first observed in central production in
p p collisions at $\sqrt{s} \ = \ 63 GeV$ by the AFS collaboration
\cite{AFS}

\begin{figure}
\epsfxsize200pt
\figurebox{}{}{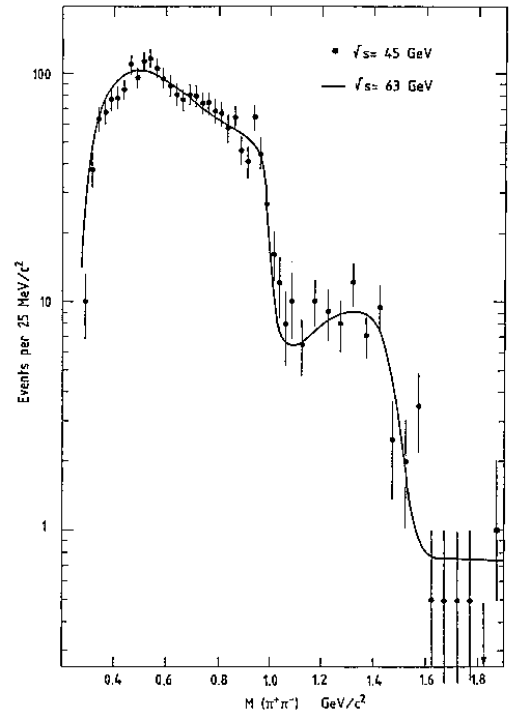}
\caption{
Invariant mass distribution of pion pairs centrally produced in pp collisions at ISR energies
\protect\cite{AFS}
}
\label{fig:AFS}
\end{figure}

\item a2) $K \overline{K}$ invariant mass distribution :

The high mass tail of $gb \ ( \ 0^{++} \ )$ as well as a separate peak from
production of $gb \ ( \ 2^{++} \ )$ can be observed in the invariant mass
distribution of $K \overline{K}$ pairs.

\end{description}

\noindent
We further propose to look for similar invariant mass distribution in multijet
dominated collisions \cite{PMWOjet}.

\noindent 

\section*{Conclusions}

\noindent
We have presented search strategies for dominant production
of binary gluonic mesons, mainly $gb \ ( \ 0^{++} \ )$ and to a lesser extent 
$gb \ ( \ 2^{++} \ )$ in high energy and high initial energy density
heavy ion as well as hadronic collisions, whence a transition through the equilibrated
quark gluon phase takes place. If these strategies prove successful,
dominant glueball production may consitute a direct and 
clear signature of the quark gluon phase
and its hadronic transition.


\begin{thebibliography}{99}

\bibitem{HFPM} H. Fritzsch and P. Minkowski,
Nuovo Cim. 30A (1975) 393.


\bibitem{PMWO} P. Minkowski and W. Ochs, 
European Physical Journal C9 (1999) 283, hep-ph/9811518.

\bibitem{SKPM} S. Kabana and P. Minkowski, 
	       Phys.Lett. B472 (2000) 155-160, hep-ph/9907570,
               S. Kabana and P. Minkowski, hep-ph/9909351.

\bibitem{Sonja1} S. Kabana, hep-ph-0004138.

\bibitem{Sonja2} S. Kabana, hep-ph/0010228, contribution to SQM2000 and Univ. of Bern 
preprint BUHE-00-06.

\bibitem{AFS} AFS Coll., T. Akesson et al., Nucl. Phys.  B264 (1986) 154.

\bibitem{SKPMn} S. Kabana and P. Minkowski, hep-ph/0010247.

\bibitem{PMWOjet} P. Minkowski and W. Ochs, Phys.Lett. B485 (2000) 139, hep-ph/0003125.


\end{thebibliography}
\end{document}